 \documentstyle[11pt,aaspp4]{article}

 \slugcomment{Accepted for publication in {\it The Astrophysical Journal\/}
9/16/96}

\lefthead{Mauche, Lee, \& Kallman}
\righthead{Ultraviolet Line Ratios}

\newcommand\lax{{\lower0.75ex\hbox{$<$}\atop\raise0.5ex\hbox{$\sim$}}}
\newcommand\gax{{\lower0.75ex\hbox{$>$}\atop\raise0.5ex\hbox{$\sim$}}}

\newdimen\minuswidth
\setbox0=\hbox{$-$}
\minuswidth=\wd0
\catcode`?=\active
\newcommand?{\kern\minuswidth}

\newdimen\digitwidth
\setbox1=\hbox{0}
\digitwidth=\wd1
\catcode`*=\active
\newcommand*{\kern\digitwidth}

\begin{document}

\title{Ultraviolet Emission Line Ratios of Cataclysmic
       Variables\altaffilmark{1}}

\author{Christopher W.\ Mauche}
\affil{Lawrence Livermore National Laboratory,\\
       L-41, P.O.\ Box 808, Livermore, CA 94550;\\
       mauche@cygnus.llnl.gov}

\author{Y.\ Paul Lee\altaffilmark{2}}
\affil{Lawrence Livermore National Laboratory, Livermore, CA 94550 and\\
       Department of Physics, University of California, Davis, CA 95616}

\and

\author{Timothy R.\ Kallman}
\affil{Laboratory for High Energy Astrophysics,\\
       NASA/Goddard Space Flight Center, Code 665, Greenbelt, MD 20771;\\
       tim@xstar.gsfc.nasa.gov}

\altaffiltext{1}{Based on observations with the {\it International Ultraviolet
Explorer\/} satellite, which is sponsored and operated by the National
Aeronautics and Space Administration, the Science Research Council of the
United Kingdom, and the European Space Agency.}

\altaffiltext{2}{Present address: Space Telescope Science Institute, 3700
                 San Martin Drive, Baltimore, MD 21218; plee@stsci.edu} 

\clearpage    


\begin{abstract}
We present a statistical analysis of the ultraviolet emission lines of 
cataclysmic variables (CVs) based on $\approx 430$ ultraviolet spectra of 20
sources extracted from the {\it International Ultraviolet Explorer\/} Uniform
Low Dispersion Archive. These spectra are used to measure the emission line
fluxes of N~V, Si~IV, C~IV, and He~II and to construct diagnostic flux ratio
diagrams. We investigate the flux ratio parameter space populated by individual
CVs and by various CV subclasses (e.g., AM~Her stars, DQ~Her stars, dwarf novae,
nova-like variables). For most systems, these ratios are clustered within a
range of $\sim 1$ decade for log Si~IV/C~IV $\approx -0.5$ and log He~II/C~IV
$\approx -1.0$ and $\sim 1.5$ decades for log N~V/C~IV $\approx -0.25$. These
ratios are compared to photoionization and collisional ionization models to
constrain the excitation mechanism and the physical conditions of the
line-emitting gas. We find that the collisional models do the poorest job of
reproducing the data. The photoionization models reproduce the Si~IV/C~IV line
ratios for some shapes of the ionizing spectrum, but the predicted N~V/C~IV
line ratios are simultaneously too low by typically $\sim 0.5$ decades.
Worse, for no parameters are any of the models able to reproduce the observed
He~II/C~IV line ratios; this ratio is far too small in the collisional and
scattering models and too large by typically $\sim 0.5$ decades in the
photoionization models.
\end{abstract}


\clearpage    


\section{Introduction}

During its 17 plus years of operation, the {\it International Ultraviolet
Explorer\/} ({\it IUE\/}) satellite (\cite{bog78}) acquired a large number
($> 10^5$) of UV spectra from all classes of astronomical objects. Cataclysmic
variables have been studied intensively with {\it IUE\/}, and these spectra
have revealed a wide variety of phenomena which were previously unknown or
poorly studied. These include mass loss via high-velocity winds, the heating
and cooling of the white dwarf, the spectrum of the accretion disk, and the
response of the disk to dwarf nova outbursts. As a result, we now have a much
more complete understanding of the nature of CVs: the geometry, velocity law,
mass-loss rate, and ionization structure of the wind; the disk instability
mechanism and the heating and cooling waves which drive the disk to and from
outburst; the relative energetics of the accretion disk and boundary layer;
subclasses; and even evolutionary scenarios. These advances are due in part
to the fact that CVs emit a large fraction of their luminosity in the UV.

With few exceptions, the strategy which has prevailed to date in the study
of CVs with {\it IUE\/} and other instruments is to make an observation or
set of observations of a single object and to analyze and interpret those
data separately from the data of other similar objects. Only when this step
is complete are the data compared with those from other similar objects. An
alternative strategy is to examine the data from all the objects of a given
class simultaneously. This is practical only in situations where a large body
of data exists, of fairly uniform quality, and when the various members of
a given class form a relatively homogeneous group. A few such studies of CVs
using various manifestations of the {\it IUE\/} archive have appeared in
the literature. Verbunt (1987) studied the absolute continuum spectral flux
distributions of disk-fed CVs. la Dous (1991) studied the relative continuum
spectral flux distributions of all classes of CVs, but concentrated on disk-fed
systems. For the discrete spectral features, only the inclination dependence of
the equivalent widths was discussed. Deng et~al.\ (1994) studied the dependence
on the orbital period and inclination of the relative continuum spectral
flux distributions and the equivalent widths of the lines of dwarf novae in
quiescence.

It is unfortunate that so little quantitative work has been performed to date
on the emission lines of CVs. Emission lines are powerful diagnostics of the
physical conditions in all types of objects including these interacting
binaries. The emission line fluxes or relative fluxes can constrain the
temperature, ionization state, density, elemental abundance,  and geometrical
distribution of the gas. However, the dependence of measurable quantities on
these parameters is often complicated, necessitating numerical model fitting
in many situations. The range of possible assumptions regarding  the physical
mechanisms of heating and ionization, together with the technical challenges
of constructing models, add to the difficulty of performing such studies. As a
result, emission lines have received less attention as diagnostics of CVs than
have continuum properties.

In the present paper, we begin to remedy this situation with an analysis of
the flux ratios of the emission lines of all classes of CVs using spectra
extracted from the {\it IUE\/} Uniform Low Dispersion Archive (ULDA). We begin
by describing the data and our analysis procedures (\S 2). We then present the
results of the data analysis (\S 3), followed by a discussion of models for
the line emission (\S 4). In \S 5 we compare data and models and in \S 6 close
with a summary of our conclusions.

\section{Data Analysis}

The {\it IUE\/} ULDA (\cite{wam89}) provides a unique source of archival data
which satisfies the requirements for a comprehensive study of a large sample of
objects owing to its uniform quality and the large number of spectra accumulated
over the long life of {\it IUE\/}. Although the {\it IUE\/} final archive
(\cite{nic94}) has since superseded the ULDA, the reprocessing effort of the
final archive had not begun when this study was begun in 1992. To build the
data sample, the {\it IUE\/} Merged Log was searched using software at the
{\it IUE\/} Regional Analysis Facility, then at the University of Colorado.
The search criteria were the camera (SWP), dispersion (low), and {\it IUE\/}
object class (dwarf novae, classic novae, irregular variables, and nova-like
variables). All spectra meeting these criteria were extracted from the ULDA
resident at Goddard Space Flight Center, which at that time contained (nearly)
all low-dispersion {\it IUE\/} spectra through 1988 (SWP sequence numbers
through $\approx 35190$). The spectra were then written to disk with header
information extracted from the {\it IUE\/} Merged Log. Excluded from further
consideration were 13 very early (prior to 1978 July) spectra of AM~Her and
SS~Cyg lacking data quality flags. The final dataset consisted of $\approx
1300$ spectra of $\approx 100$ CVs.

From this dataset, we selected all spectra exhibiting pure emission lines.
Sources which met this criteria include AM~Her stars, DQ~Her stars, dwarf
novae in quiescence, and eclipsing nova-like variables and dwarf novae in
outburst. To this dataset we added the {\it IUE\/} spectra of the eclipsing
nova-like variable V374~Pup listed in Table~1 of Mauche et~al.\ (1994), as
we were working on that source at the time. Excluded from analysis were
non-eclipsing nova-like variables and dwarf novae in outburst, since the UV
resonance lines of such systems are either in absorption or are P~Cygni
profiles.

For these spectra, we measured the integrated fluxes above the continuum of
the strongest emission lines in the SWP bandpass: N~V $\lambda 1240$, Si~IV
$\lambda 1400$, C~IV $\lambda 1550$, and He~II $\lambda 1640$. To determine
the flux of these lines, the continuum to the left and right of each line
(e.g., from 1500--1530 \AA \ and from 1570--1600 \AA \ for C~IV) was fitted
with a linear function by the method of least squares. This method works well
with all of the emission lines except N~V, because the left side of this line
falls on the wing of the geocoronal Lyman $\alpha $ emission line. To overcome
this problem, a continuum point to the left of the N~V line was specified
interactively, and the fit included the region to the right of the line
(1255--1270 \AA ) plus this continuum point to the left. Care was taken to
insure that the continuum over this small wavelength region was a reasonable
extrapolation both in normalization and slope to the continuum at longer
wavelengths. In all cases the least-squares fit was made under the assumption
that the errors $\sigma _i$ in the flux density $f_i$ at each wavelength point
$\lambda _i$ were the same. After the normalization $a$ and slope $b$ of the
continuum was determined, the size of the errors $\sigma $ were estimated by
assuming that the reduced $\chi ^2$ of the fit was equal to 1: $\sigma ^2 =
\sum _{i=1}^N (f_i-a-b\lambda _i)^2/(N-2)$. This procedure is necessary
because the ULDA (unlike the {\it IUE\/} final archive) does not include the
error associated with each flux point. The integrated fluxes of the emission
lines were then determined by summing the flux minus the fitted continuum
over the region of the line (e.g., 1530--1570 \AA \ for C~IV). The error on
the integrated flux was determined by propagating the error $\sigma $ through
the calculation. Finally, careful track was kept of the number of pixels
labeled by the data quality flags as either saturated or nearly saturated
(extrapolated ITF). Such pixels compromise the quality of the flux density
measurements and consequently the integrated line fluxes; saturated pixels
are particularly bad in this regard, because they systematically produce
only lower limits to the flux density. In the final cut to the data, flux
measurements for a given line were retained only if the total number of
saturated and extrapolated pixels included in the summation was less than
or equal to some number $n$. While $n=1$ would have been the ideal choice,
such a stringent selection criterion would cull too many measurements. As
a reasonable compromise, we settled on $n=3$ for this selection criterion,
allowing a few mild sinners to pass with the saints. The roster at this stage
of the analysis consisted of $\approx 700$ spectra of $\approx 60$ CVs. Of
this number, we present results based on $\approx 430$ spectra of the 20
systems with the largest number ($> 10$) of flux measurements.

\section{Observational Results}

Just as color-color diagrams are a useful way to discuss the continuum
spectral flux distributions of CVs (e.g., \cite{wad88}; \cite{den94}), flux
ratio diagrams are a useful way to display and discuss the fluxes of their
emission lines. Flux ratios remove the variables of luminosity and distance
from the comparison of different sources; flux ratios are unaffected by
aperture losses, which occur for {\it IUE\/} spectra obtained through the
small aperture; and flux ratios are much less sensitive to reddening than the
fluxes themselves. Due to the increase of the extinction efficiency at short
wavelengths, reddening has the largest differential effect on the fluxes of
the He~II and N~V lines. Luckily, for a typical extinction of $E_{B-V}\leq
0.1$ (\cite{ver87}), the ratio of the measured fluxes of these lines is only
$\leq 18$\% higher than intrinsic ratio. As we shall see, this correction,
while systematic, is much smaller than either the typical measurement errors,
or the typical dispersion in the flux ratios of any given source.

Of the many possibilities, we choose to plot the ratios N~V $\lambda
1240$/C~IV $\lambda 1550$ and He~II $\lambda 1640$/C~IV $\lambda 1550$ vs.\
Si~IV $\lambda 1400$/C~IV $\lambda 1550$ (Figs.~1--8). In what follows, we 
often refer to these ratios simply as N~V/C~IV and He~II/C~IV vs.\ Si~IV/C~IV.
It should be emphasized that the identifications of these lines are not unique;
at $\Delta\lambda\approx 5$~\AA , the {\it IUE\/} resolution is not sufficient
to exclude several other possible lines in the vicinity of these wavelengths.  
We defer a discussion of such potential confusion, and of the physical
motivation for our choice of lines  until the following section, and begin by
presenting the phenomenological behavior of these ratios as derived from the
data. We begin by discussing the various members of the various CV subclasses
and then discuss the sources which do not appear to fit the pattern established
by the other members of their subclass (sources we enjoy referring to as
``weird'').


\subsection{Results Arranged by Subclass}

\subsubsection{AM~Her stars: AM~Her, V834~Cen, and QQ~Vul}

Figure~1 shows that AM~Her, V834~Cen, and QQ~Vul form a sequence of increasing
N~V/C~IV with increasing Si~IV/C~IV. He~II/C~IV is reasonable constant at
$\approx 0.2\pm 0.1$.

\subsubsection{DQ~Her stars: EX~Hya, TV~Col, FO~Aqr, and DQ~Her}

Among these DQ~Her stars, TV~Col displays the greatest amount of variability
in its continuum and line fluxes, and, as is evident in Figure~2, in its UV
line ratios. The upper right portion of the observed range of line ratios
is populated by the observations obtained during the optical and UV flare
observed by Szkody \& Mateo (1984); during the flare, both N~V and He~II
increased relative to C~IV. The relatively tight phase space occupied by the
line ratios of FO~Aqr is approximately the same as that of TV~Col in outburst.
The ionization state of the line-emitting gas in EX~Hya and DQ~Her is lower
than the two other DQ~Hers, as evidenced by their higher Si~IV/C~IV line
ratios and the lower He~II/C~IV line ratios.

\subsubsection{Dwarf Novae in Quiescence: SS~Cyg, SU~UMa, RU~Peg, and WX~Hya}

As shown in Figure~3, WX~Hyi has the largest ratios of N~V/C~IV and Si~IV/C~IV
among all the AM~Her stars, DQ~Her stars, and other dwarf novae in quiescence.
SU~UMa and RU~Peg have N~V/C~IV ratios of $\approx 0.3$, but the value for
SS~Cyg is $\approx 30\%$ lower at $\approx 0.2$.

\subsubsection{Eclipsing Dwarf Novae: Z~Cha and OY~Car}

As shown in Figure~4, relative to all other ``normal'' CVs, Z~Cha and OY~Car
both have strong N~V relative to C~IV. Almost all of the spectra of Z~Cha
were obtained during the peak and decline of the superoutburst of 1987 April
(\cite{har2a}) and of a normal outburst of 1988 January (\cite{har2b}).
The evolution of the spectrum during these observations is manifest most
clearly in the Si~IV/C~IV ratio, which was highest during the decline of the
superoutburst ($V\approx 12.3$--12.7) and lowest during the decline of the
normal outburst ($V\approx 13.3$--14.0). During these intervals, both the
N~V/C~IV and He~II/C~IV ratios remained roughly constant, though there is
some indication that the He~II/C~IV ratio increased as the Si~IV/C~IV ratio
decreased. The spectra of OY~Car were obtained during the decline of the
superoutburst of 1985 May (\cite{nay88}) and produce flux ratios which
roughly equal those of Z~Cha in superoutburst.

\subsubsection{Eclipsing Nova-like Variables: UX~UMa, V347~Pup, and RW~Tri}

As shown in Figure~5, the eclipsing nova-like variables UX~UMa, V347~Pup,
and RW~Tri have moderately strong N~V relative to C~IV, with N~V/C~IV ratios
intermediate between those of dwarf novae in quiescence and eclipsing dwarf
novae. The phase space occupied by V347~Pup is amazingly tight, and is disturbed
only  by eclipse effects: the single discrepant Si~IV/C~IV ratio was obtained
from the eclipse spectrum of this source, and is due to that fact that the
C~IV emission line is eclipsed less than the other emission lines. The cause
of the large spread in line ratios of UX~UMa appears to due to a different
mechanism: there are epochs when Si~IV is weak relative to C~IV, and other
epochs when it is reasonably strong. These epochs do not seem to correlate
with the continuum flux. The He~II/Si~IV ratio of UX~UMa is typically less
than that of V347~Pup.

\subsubsection{Intercomparison of Various Subclasses}

Figure~6 combines the line ratios of the objects from the previous figures
grouped together into magnetic (upper panels) and non-magnetic (lower panels)
systems. It is apparent from this figure and the previous discussion that the
line ratios of the various subclasses are rather homogeneous, with a dispersion
of $\sim 1$ decade. This dispersion to be due to almost equal contributions
from the dispersion of values from the various observations of individual
objects ($\sim 0.5$ decade) and from the dispersion from one object to another
($\sim 0.5$ decade).

There are clear and significant differences between these two subclasses: (i)
He~II/C~IV is $\sim 0.25$ decades larger in magnetic systems; (ii) Si~IV/C~IV
is slightly larger in the non-magnetic systems, but the difference is small
compared to the dispersion; (iii) The upper limit on the N~V/C~IV distribution
is greater in the non-magnetic systems, while the lower limit of the
distribution is similar in the two cases.

It is also instructive to consider the differences within these two subclasses.
The greater range of the N~V/C~IV distribution is due to the eclipsing dwarf
novae Z~Cha and OY~Car (squares and circles, respectively, in the lower panels
of Fig.~6); otherwise the N~V/C~IV distributions of magnetic and non-magnetic
systems are similar. Among the magnetic systems, TV Col (triangles in the upper
panels of Fig.~6) has significantly lower Si~IV/C~IV than any other object; the
Si~IV/C~IV distribution would be much tighter if this object were excluded from
the sample.

\subsection{Systems Showing Anomalous Behavior: ``Weird'' CVs}

\subsubsection{GK Per}

GK~Per is unusual among CVs in a number of respects. It has an evolved secondary
and hence a very long orbital period and a large accretion disk. Unlike most
CVs, the UV continuum of this DQ~Her-type system peaks in the {\it IUE\/}
bandpass. The cause of this anomaly is thought to be due to the disruption
of the inner disk by the magnetic field of the white dwarf (\cite{bia83};
\cite{wu89}; \cite{mau90}; \cite{kim92}). As shown in Figure~7, GK~Per also
distinguishes itself from other CVs in its  anomalously large He~II/C~IV ratio.
In outburst, both He~II and N~V are stronger than C~IV, whereas the opposite is
true in quiescence. While the Si~IV/C~IV ratio spans a rather broad range of
$\approx 0.2$--1, this range is present in both outburst and quiescence.

\subsubsection{V~Sge}

V~Sge is an unusual nova-like variable which occasionally manifests
brightenings of as much as 3 mag (\cite{her65}). The model for this system is
highly uncertain; both white dwarf and neutron star binary models have been
considered (\cite{koc86}; \cite{wil86}). The {\it IUE\/} spectra of V~Sge
obtained during 1978 and 1979 are characterized by strong He~II and N~V emission
lines relative to C~IV (\cite{koc86}): N~V/C~IV $\approx 2.5$ and He~II/C~IV
$\approx 3$. These distinguishing line ratios are suppressed in outburst. In
1985 August, the UV continuum was enhanced by factor of $\approx 2$ relative
to the earlier spectra, while the N~V/C~IV ratio fell to $\approx 0.4$ and the
He~II/C~IV ratio fell to $\approx 1$. In 1985 April, the UV continuum was
enhanced by factor of $\approx 4$, while the N~V/C~IV ratio fell to $\approx
0.15$; measurements of the He~II/C~IV ratio are not possible during this epoch
because the He~II emission line was overexposed. This behavior is opposite to
that of GK~Per, whose N~V/C~IV and  He~II/C~IV ratios increased mildly in
outburst.

\subsubsection{BY~Cam}

BY~Cam (H~0538+608) is one of only three AM~Her stars above the period gap
(V1500 Cyg and AM~Her being the other two) and one of only two which rotates
asynchronously (V1500 Cyg being the other one). As pointed out by
Bonnet-Bidaud \& Mouchet (1987), BY~Cam is unusual in having weak C~IV and
strong N~V. Figure~7 shows that all the line ratios are large because C~IV
is weak.

\subsubsection{AE~Aqr}

AE~Aqr exhibits a host of bizarre phenomena including variable flare-like
radio and optical emission; rapid, coherent oscillations; and QPOs. Its UV
spectrum has been discussed by Jameson et~al.\ (1980) and is distinguished
by the near-absence of C~IV. Unfortunately, the flux measurements of the C~IV
emission line are very uncertain because this line is so weak and because the
measurements are possibly contaminated by the Si~II 3p--4s multiplet at 1526.71,
1533.43~\AA \ (e.g., \cite{kel87}). Nevertheless, it is clear that N~V and
Si~IV have similar strengths and are $\approx 10$ times stronger than He~II
and C~IV, which have similar strengths. 

\section{Line Ratio Modeling}

We now present some simple models for the systematic behavior we expect from CV
line ratios. Although CVs as a class share a number of fundamental properties,
they also differ between subclasses in the likely properties of the gas
responsible for line emission. For example, in disk-fed systems, such as
nova-like variables and dwarf novae, a possible site for line emission is the
disk atmosphere. In this case, the atmosphere may be heated either by viscosity
(e.g., \cite{sha91}) or by photons from other parts of the disk or from the
vicinity of the white dwarf (e.g., \cite{ko96}). A wind is also present in
nova-like variables and dwarf novae in outburst (see \cite{mau6b} for a recent
review of CV winds). In AM~Her stars, the line emission may come from the stream
of material being transferred from the secondary to the white dwarf which is
heated either mechanically or by photons from the shock at the white dwarf
surface. DQ~Her stars may have emission from either or both of these sites. 

Beyond these simple considerations, there is considerable uncertainty about the
nature of the emission region: the geometrical arrangement of the gas relative
to the source of ionizing photons (if present), the density and optical depth
of the gas, the elemental abundances of the gas, the spectrum and flux of the
ionizing radiation, and the presence and distribution of any other sources of
heating. These quantities are not only uncertain, they may differ from one
object to another in a subclass. Furthermore, all subclasses contain anomalous
members. We therefore consider models for CV emission regions which employ the
very simplest assumptions, namely, slab models each with a single ionizing
spectrum, gas composition, and density. We then explore plausible choices for
photon flux, column density, and mechanical heating rate (if any). The results
serve not only as a test of the validity of the models when compared to the
observations, but also provide insight into the physics of the line emission
which will hopefully remain useful when more detailed models are  considered. 

Further support for this strategy comes from the results of the previous
section: the objects of a given subclass show significant clustering in the
two line ratio diagrams, and there are clear differences from one subclass
to another. Therefore, we can define our modeling goals as follows: to test
whether one or more sets of models can reproduce the mean values of the observed
line ratios for the various subclasses of objects, and to test whether any
models can reproduce the dispersion or systematic variability of the observed
line ratios. Given the uncertainties of the models and the observed dispersion
in line ratios, we will consider agreement between the models and observations
to be adequate if the two agree to within the dispersion of the observed values,
i.e., approximately one decade. We will show that although this is clearly a
very crude criterion, it turns out to be very constraining for the models.

\subsection{Model Ingredients}

Although the mechanism of line emission is uncertain, we favor photoionization
as the source of ionization, excitation, and heating of the gas. This is partly
because the observed continuum spectra, and their extrapolation into the
unobservable spectral regions, provide a convenient and plausible energy source.
Further support for this idea comes from Jameson et~al.\ (1980) and King et~al.\
(1983) who compared the observed line strengths from AE Aqr and UX UMa to
simple predictions of both photoionized and coronal models and found the coronal
models inconsistent with the observed lines. Nevertheless, for the sake of
completeness, we examine both photoionized and coronal models.

The models are calculated using the XSTAR v1.19 photoionization code
(\cite{kal82}; \cite{kal93}). The models consist of a spherical shell of gas
with a point source of continuum radiation at the center; this may be used to
represent a slab in the limit that the shell thickness is small compared with
the radius. In what follows we will use the line fluxes emergent from the
illuminated face of the slab, which is equivalent to those emitted into the
interior of the spherical shell. The input parameters include the source
spectrum, the gas composition and density, the initial ionization parameter
(determining the initial radius, see below for a definition), and the column
density of the shell (determining the outer radius). Construction of a model
consists of the simultaneous determination of the state of the gas and the
radiation field as a function of distance from the source. The state of the
gas at each radius follows from the assumption of a stationary local balance
between heating and cooling and between ionization and recombination.

When the gas is optically thin, the radiation field at each radius is determined
simply by geometrical dilution of the given source spectrum. Then, as shown
by Tarter et~al.\ (1969), the state of the gas depends only on the ionization
parameter $\xi$, which is proportional to the ratio of the radiation flux to the
gas density. We adopt the definition of the ionization parameter used by Tarter 
et~al.: $\xi=L/(nr^2)$, where $L$ is the ionizing energy luminosity of the
central source (between 1 and 1000 Ry), $n$ is the gas density, and $r$ is the
distance from the source. This scaling law allows the results of one model
calculation to be applied to a wide variety of situations. For a given choice of
spectral shape, this parameter is proportional to the various other customary
ionization parameter definitions: $U_H=F_H/n$, where $F_H$ is the incident
photon number flux above the hydrogen Lyman limit; to $\Gamma=
F_\nu(\nu_L)/(2hcn)$, where $F_\nu(\nu_L)$ is incident energy flux at the Lyman
limit; and to $\Xi=L/(4\pi R^2 cnkT)$. This simple picture breaks down when the
cloud optical depth is non-negligible, since the source spectrum then depends
on position, and the escape of cooling radiation in lines and recombination
continua depends on the total column density of each ion species and hence on
the ionization state of the gas throughout the cloud. In addition, the rates for
cooling and line emission can depend on gas density owing to the density
dependence of line collisional deexcitation and dielectronic recombination.
However, even in this case the ionization parameter remains a convenient means
of characterizing the results.

The state of the gas is defined by its temperature and by the ion abundances. 
All ions are predominantly in the ground state, and except for hydrogen and
helium the populations of excited levels may be neglected. The relative
abundances of the ions of a given element are found by solving the ionization
equilibrium equations under the assumption of local balance, subject to the
constraint of particle number conservation for each element. Ionization balance
is affected by a variety of physical processes, most notably photoionization
and radiative and dielectronic recombination. The temperature is found by
solving the equation of thermal equilibrium, by equating the net heating of the
gas due to absorption of incident radiation with cooling due to emission by
the gas. These rates are derived from integrals over the absorbed and emitted
radiation spectra. Although Compton scattering is not explicitly included as a
source or sink of radiation, its effect is included in the calculation of the
thermal balance.

The emitted spectrum includes continuum emission by bremsstrahlung and 
recombination and line emission by a variety of processes including
recombination, collisions, and fluorescence following inner shell 
photoionization. Line transfer is treated using an escape probability
formalism and includes the effects of line destruction by collisions and 
continuum absorption. Transfer of the continuum is calculated using a
single stream approximation, as described in Kallman \& McCray (1982).

Rates for atomic processes involving electron collisions have been modified
since the publication of Kallman \& McCray (1982) to be consistent with
those used by Raymond \& Smith (1977). Recombination and ionization rates for
Fe have been updated to those of Arnaud \& Raymond (1992). In addition, we
have added many optical and UV lines from ions of medium-Z elements (C, N, O,
Ne, Si, and S) using collisional and radiative rates from Mendoza (1983). The
elements Mg, Ar, Ca, and Ni have also been added. The models have a total of
168 ions, producing 1715 lines, of which 665 have energies greater than 120 eV
(10 \AA ), and approximately 800 are resonance lines. For each ion we also
calculate the emission from radiative recombination onto all the excited levels
which produce resonance lines. The number of such continua is equal to the
number of resonance lines in the calculation. Some of the results of the models
are sensitive to the rates for dielectronic recombination. We use the high
temperature rates given by Aldrovandi \& Pequignot (1973), together with the
low temperature rates from Nussbaumer \& Storey (1983). These rates differ
significantly for several relevant ions, e.g., Si~III, relative to those of
Shull \& Van Steenberg (1983), which were used in most earlier versions of
XSTAR. Finally, all models assume element abundances which are close to solar:
H:He:C:N:O:Ne:Mg:Si:S:Ar:Ca:Fe:Ni =
1:0.1:3.7E-4:1.1E-4:6.8E-4:2.8E-5:3.5E-5:3.5E-5:1.6E-5:4.5E-06:2.1E-6:2.5E-5:2.0E-6
(\cite{wit71}).

An emission line may be produced by two types of physical processes. First is
what we will call ``thermal emission,'' which is recombination or collisional
excitation by thermal electrons. Second is excitation by photons from the
incident continuum. Since this is an elastic scattering process, we will refer
to it as ``scattering.'' This process will produce an apparent emission feature
if our line of sight to the continuum source is at least partially blocked, and
the scattering region must be concentrated on the plane of the sky (see
\cite{kro95} for more discussion of these issues). If the scattering region has
a bulk velocity greater than the thermal line width then a P~Cygni or inverse
P~Cygni profile will form even if the scattering region is spherical. Scattering
fails to account for the presence of He~II $\lambda$1640, since this is a
subordinate line and the population of the lower level will be negligibly small
under conditions appropriate to photoionized gas. Significant opacity in this
line would require a temperature greater than $\sim 3\times 10^5$ K and
level populations which are close to LTE values. In spite of this, we have
investigated the possibility that this mechanism can explain the ratios of the
other lines we consider. We assume that the fluxes in scattered lines are
proportional to the opacities in the lines. This is likely to be justified if
the line optical depths are less than unity, and if the wind ionization balance
is approximately uniform. If so, the line flux ratios will equal the opacity
ratios.

\subsection{Input Parameters}

The models we explore fall into two categories: photoionized models and
collisional (coronal) models. For each we present the line ratios for both
thermal emission and scattering emission mechanisms. The photoionization model
parameters are chosen in an attempt to crudely represent the range of observed
ionizing spectra from CVs. These typically consist of a soft component which is
consistent with a 10--50 eV blackbody, together with a hard X-ray component such
as a 10 keV bremsstrahlung (e.g., \cite{cor2a}). The ratio of strengths of these
two components varies from one object to another, and with outburst state and
subclass. However, a typical ratio is 100:1 (soft:hard) for non-magnetic systems,
and 1:1 for magnetic systems (\cite{cor2b}). Beyond such simple considerations,
the detailed shape of the ionizing radiation field is difficult to determine
accurately. This is due to the strong influence of photoelectric absorption by
interstellar and circumstellar gas, and to the limited bandpass and spectral
resolution of most past observations in the soft X-ray band (e.g., \cite{ram94};
\cite{mau6a}). Owing to such uncertainties we choose a few very simple ionizing
spectra for consideration in our photoionization models: Model A: 30 eV
blackbody, Model B: 10 keV brems, Model C: 50 eV blackbody, Model D: 10 eV
blackbody. Model G is the mechanically heated (coronal) ionization model. All
these models have a total (neutral + ionized) hydrogen column density of $N_{\rm
H}=10^{19}~\rm cm^{-2}$, chosen to make them optically thin to the continuum and
effectively thin to the escaping resonance lines. In addition, we present two
photoionized models which are close to being optically thick, both of which have
of column density $N_{\rm H}=10^{23}~\rm cm^{-2}$. Model E has a 30 eV blackbody
ionizing spectrum, and Model F has a 10 keV brems ionizing spectrum. For each
model we determine the net line flux contained in the wavelength intervals
1237--1243~\AA , 1390--1403~\AA , 1547--1551~\AA , and 1639--1641~\AA , which we
refer to as N~V, Si~IV, C~IV, and He~II, respectively. As we will show, these
wavelength intervals also contain other lines which can mimic these strong
lines, and these other lines are likely to affect the interpretation of the {\it
IUE\/} low resolution data as well. The results of our models---the Si~IV/C~IV,
N~V/C~IV, and He~II/C~IV line ratios---are summarized in Tables 1--3. For the
photoionization models, we consider various values of the ionization parameter
for six choices of ionizing continuum shape. For the collisional models, we
consider ten values of the gas temperature. We also tested for the dependence on
gas density and found that it is negligible for the conditions we consider; all
the models presented here have density $n=10^9~\rm cm^{-3}$.

\subsection{Model Results}

When the line ratios shown in Tables 1--3 are plotted in two diagrams in the
same way as the {\it IUE\/} data, several common properties emerge. These are
shown in Figures 9 and 10, with various symbols denoting the ionizing spectra:
Model A \vrule height6pt width5pt depth-1pt, Model B = $\bullet $, Model C =
$\times $, Model D = +, Model E = $\Box $, Model F = $\circ $. We have also
tried a model consisting of a 30 eV blackbody together with a 10 keV brems
spectrum in a ratio of 99:1; the results are so similar to Model A as to be
indistinguishable, i.e., the 30 eV blackbody has far more influence on the
ionization balance than does the 10 keV brems in these ratios. Model G, the
coronal case, produces line ratio combinations which are almost entirely outside
the range spanned by these figures. For the other spectra, as expected, there is
little or no dependence on model density for optically thin photoionized models.

The shape of the trajectory of the model results in these planes is similar
for most of the models. They resemble a {\sf U} shape, although in some cases
the right upright of the {\sf U} is missing, and in others it is tipped nearly
45 degrees to the vertical. We can understand the results better if we label
the points along the {\sf U} as follows: A = upper left extreme of trajectory;
B = bottom of steepest part of left upright; C = lowest point of trajectory; D =
bottom of right upright; E = upper right extreme of trajectory. These are shown
schematically in Figure~11. In general, the trajectory is traversed from point A
to point E as the ionization parameter decreases, and may be understood in terms
of the relative ease of ionization of the various ions responsible for line
emission. The ionization parameter at which the abundance of a given ion peaks,
relative to its parent element, can be derived crudely from the ionization
potential. Thus, the four ions responsible for the strongest observed lines may
be ordered in terms of decreasing ease of ionization according to: Si~IV, He~II,
C~IV, N~V. For the metal ions the ionization parameter at which the emissivities
of the lines peak is approximately the same as the ionization parameter at which
the ion abundances peak. Thus, at the highest ionization parameters we consider,
nitrogen is ionized to or beyond N~V, carbon is ionized beyond C~IV, and silicon
is ionized beyond Si~IV. However, the abundance of N~V, and hence its line
emissivity, is greater than the corresponding quantities for C~IV, which in
turn are greater than for Si~IV. Thus, at high ionization parameter, N~V/C~IV
is relatively large and Si~IV/C~IV is small, corresponding to point A of the
trajectory in the N~V/C~IV vs.\ Si~IV/C~IV plane. At intermediate ionization
parameter, carbon recombines to C~IV and nitrogen recombines below N~V, so that
N V/C~IV is small and Si~IV/C~IV is also small (point B). At lower ionization
parameter, silicon recombines to Si~IV and carbon may recombine to below C~IV,
resulting in an intermediate value of Si~IV/C~IV (point C). At very low
ionization parameter, Si~IV/C~IV is at a maximum, and there is an apparent
increase in N~V/C~IV (points D, E). This cannot be understood in terms of
conventional ionization balance, since the C~IV abundance always exceeds the N~V
abundance at low ionization parameter. Rather, it is due to confusion between
the N~V 2s--2p doublet at 1238.82, 1242.80~\AA \ and the Mg~II 3s--4p doublet at
1239.93, 1240.39~\AA \ (e.g., \cite{kel87}). Since Mg~II has a lower ionization 
potential than any of the other ions in question, Mg~II/C~IV increases at low
ionization parameter, thus explaining the apparent increase in N~V/C~IV .

The behavior of He~II $\lambda$1640 differs from the other lines owing to
the fact that it is emitted by recombination, while the others are emitted by
collisional excitation. Thus, the emissivity of He~II $\lambda$1640 remains
nearly constant at high ionization parameter, while the C~IV $\lambda$1550 line
emissivity decreases with increasing ionization parameter. This is in spite of
the fact that He~II is more easily ionized than C~IV. This explains the AB
part of the trajectory in the He~II/C~IV vs.\ Si~IV/C~IV plane. As the
ionization parameter decreases, C~IV recombines to C~III and below, while He~II
and He~III are still abundant, thus explaining the DE part of the trajectory.

The model behavior under the scattering scenario is qualitatively similar to 
the thermal excitation scenario, except for the weakness of He~II. The line 
strengths in the scattering case are less dependent on the gas temperature
than in the thermal excitation case, and the oscillator strength for the Mg~II
3s--4p transition is small, so that the trajectory lacks the DE segment in the
N~V/C~IV vs.\ Si~IV/C~IV plane.

The fact that the coronal models fail almost completely to produce line ratios 
within the range of our diagrams indicates that it is unlikely that this
process dominates in CV line-emitting gas. This result is not surprising, owing
to the well-known fact that coronal equilibrium produces ion abundances that
have less overlap in parameter space (e.g., temperature) between adjacent ion
stages than does photoionization.

\subsection{Spectral Dependence}

The shape of the ionizing spectrum influences the location of the various 
points along the trajectory in the two line ratio diagrams, and also the 
existence of part of the trajectory, most notably the segments between points 
C and E. 

For soft spectra, such as the 30 eV blackbody, recombination to species below
Si~IV does not occur for the parameter range considered ($\log\xi =-1.5$ to
+1.0), so the CDE part of the trajectory is absent in the N~V/C~IV vs.\
Si~IV/C~IV plane. Also, when the ionization parameter is suitable for producing
large N~V/C~IV, the Si~IV/C~IV ratio is so small as to be off the scale of
Figure~9.

For very soft spectra, such as the 10 eV blackbody, there are insufficient hard
photons to produce N~V. Thus, points A and B are missing in the N~V/C~IV vs.\
Si~IV/C~IV plane. The DE part of the trajectory is entirely due to Mg~II
$\lambda1240$.

For hard spectra, such as the 10 keV brems spectrum, X-rays can make Si~IV via
Auger ionization even at very low ionization parameter. Thus, the range of
Si~IV/C~IV is greatly expanded, and can reach 10 at point D in the N~V/C~IV
vs.\ Si~IV/C~IV plane. Other models, such as a 30 eV blackbody, can make large
Si~IV/C~IV at low ionization parameter, but they also have very weak N~V line
emission, so that the 1240~\AA \ feature is dominated by Mg~II and they are on
DE segment of the trajectory.

At high ionization parameter He~II is a ``bolometer'' of the ionizing spectrum,
since it is dominated by recombination. That is, the He~II strength depends only
on the number of photons in the He~II Lyman continuum ($\varepsilon\geq 54.4$
eV). So, harder spectra make stronger He~II, and conversely. This behavior may
still hold at point C. At small ionization parameter, He~II is likely to be more
abundant than C~IV or Si~IV. This explains the CDE segment of the trajectory
in the He~II/C~IV vs.\ Si~IV/C~IV plane. This fact does not seem to depend
strongly on spectral shape, although the value of Si~IV/C~IV at points C, D,
and E does depend on the spectrum; this ratio increases at all these points for
harder spectra, and conversely.

The 10 eV blackbody models show a ``hook'' in their trajectory in which the
Si~IV/C~IV appears to increase with increasing ionization parameter at the upper
end of the range of values we consider. This is counter to the expected behavior
of Si~IV at high ionization parameter, since we expect silicon to become ionized
beyond Si~IV and the Si~IV abundance to decrease at high ionization parameter.
The reason for the model behavior is the confusion between the Si~IV doublet
at 1393.76, 1402.77~\AA \ and the O~IV $\rm 2s^22p$--$\rm 2s2p^2$ multiplet
at 1397--1407~\AA \ (e.g., \cite{kel87}). The unique behavior of the 10 eV
blackbody models is due to the fact that the O~IV lines increase at high
ionization parameter only for the softest spectra; harder spectra ionize oxygen
past O~IV when other ion abundances are at similar values. Like the N~V and
Mg~II lines near 1240~\AA , the Si~IV and O~IV lines near 1400~\AA \ can be
confused in {\it IUE\/} low resolution data.

The overlap in ionization parameter space of the regions where N~V, Si~IV,
C~IV, and He~II predominate does not differ greatly between models with hard
X-ray spectra (e.g., 10 keV brems) and those with blackbodies with $kT>30$ eV.
A more pronounced difference is due to the fact that the latter spectra have
all their ionizing photons crammed into a smaller energy range. Therefore, the
ionization parameter scale, which simply counts the energy in ionizing photons,
and the distribution of ionization, which really depends on the photon density
in the EUV/soft X-ray region, are very different in the two cases. For example,
in the 10 keV brems case, C~IV predominates at $\log\xi =0$, while in the 30 eV
blackbody case it predominates near $\log\xi=-1$. Thus, our model grid, which
spans the range $-3< \log\xi < +1.5$, does not include the region where the gas
has recombined below C~IV, etc., in the blackbody case. This accounts for the
absence of the CDE part of the line ratio trajectory for the blackbody models.

\section{Comparison with Observations}

The models presented in the previous section provide a useful context in which
to examine the likely physical conditions in CV line-emitting regions. As was
discussed earlier, we do not expect these simple models to account for the
details of the observed spectra, but we do hope that they will at least crudely
reproduce some of the features of the observations. These might include: the
range of the observed ratios to within the dispersion of the observed values,
i.e., approximately one decade, and possibly the differential behavior of a
given source as it varies in time. In fact, we find little evidence for
agreement between the observations and any of the models beyond the simplest
measures of consistency for some of the line ratios. We begin the comparison of
model results with observations by considering the ``normal'' CVs.

\subsection{``Normal'' CVs}

The observed ratios of ``normal'' CVs lumped together regardless of class
(Fig.~6) are clustered within a range of $\sim 1$ decade for log Si~IV/C~IV
$\approx -0.5$ and log He~II/C~IV $\approx -1.0$ and $\sim 1.5$ decades for log
N~V/C~IV $\approx -0.25$. The larger range of the N~V/C~IV ratio is due largely
to the large line ratios of the eclipsing dwarf novae Z~Cha and OY~Car (squares
and circles, respectively, in Fig.~6). Otherwise, the range is $\sim 1$ decade
centered on $\approx -0.5$. This same general range is spanned by the models,
but there is little detailed agreement.

One notable failure of the models is the behavior in the He~II/C~IV vs.\
Si~IV/C~IV plane. The photoionization models always produce He~II/C~IV $\gax$
Si~IV/C~IV. This is because, although the He~II line is emitted following
radiative recombination and the C~IV and Si~IV lines are formed by collisional
excitation, photons at energies greater than 54.4 eV which are responsible
for ionizing He~II are also responsible for heating the gas. So, models which
efficiently heat the gas and emit Si~IV and C~IV also have efficient ionization
of He~II and hence efficient production of the 1640~\AA \ line. In contrast,
``normal'' CVs show He~II/C~IV values which are less than Si~IV/C~IV by $\sim
0.5$ decades. Magnetic systems have He~II/C~IV line ratios which are
systematically higher (and Si~IV/C~IV line ratios which are systematically 
lower) than those of the non-magnetic systems by $\sim 0.25$ decades. Only the
DQ~Her stars TV~Col and FO~Aqr (triangles and pentagons, respectively, in
Fig.~6) have He~II/C~IV $\geq$ Si~IV/C~IV. 

It is interesting to note that the ``hook'' in the trajectory of the 10 eV
blackbody models referred to in the previous section occurs near the values of
these line ratios where most observed objects cluster. Furthermore, the 10 eV
models come closest to reproducing the observed He~II/C~IV ratios; they are
the only ones for which He~II/C~IV falls below Si~IV/C~IV. 

The coronal models produce He~II/C~IV values of order 1\% of Si~IV/C~IV when
this latter ratio is in the observed range; even at this relatively favorable
point, the N~V/C~IV ratio is much smaller than observed. This suggests that
the coronal emission mechanism is less likely than photoionization for all CVs.

In the N~V/C~IV vs.\ Si~IV/C~IV plane, the models and the data span a similar
range of ratios, so there is less indication of the failure of any of the
models. The 50 eV blackbody models are least successful at reproducing the 
most commonly observed values of these ratios simultaneously. The 30 eV
blackbody and 10 keV brems models both span the observed range, as do the
optically thick models (which also use these ionizing spectra). The ``hook''
in the trajectory of the 10 eV blackbody models causes the Si~IV/C~IV ratio to
lie almost entirely in the range $-1.5\leq $ log Si~IV/C~IV $\leq -0.5$, which
is close to that spanned by the observed ratios of most ``normal'' CVs.

In addition to asking whether there exists a model which can reproduce a given
ratio, we can ask whether the distribution of observed ratios is consistent with
the analogous model quantity. For example, if the ionizing spectrum and emission
mechanism are independent of time, but the luminosity and hence the ionization 
parameter varies with time, we expect that the ratios of a given object will
lie along a {\sf U}-shaped trajectory in the line ratio diagram. In contrast,
there appears no clear pattern in the observed ratios for objects with many
observations, other than a clustering in a well-defined region of the diagram. 

\subsection{``Weird'' CVs}

Consider next the ``weird'' CVs. Figures~7 and 8 demonstrate that V~Sge, GK~Per,
BY~Cam, and AE~Aqr form a sequence of dramatically increasing Si~IV/C~IV and
N~V/C~IV at nearly constant He~II/C~IV. In V~Sge and GK~Per, He~II/C~IV $>$
Si~IV/C~IV, unlike most of the ``normal'' CVs, but consistent with the
photoionization models. The extreme line ratios of BY~Cam and AE~Aqr are harder
to understand. Bonnet-Bidaud \& Mouchet (1987) have suggested a depletion of
carbon by a factor of $\sim 10$ to explain the anomalous line ratios of BY~Cam.
If a similar deficiency applies to AE~Aqr, depletion by a factor of $\sim 60$
is required. Although such abundance anomalies are possible, they are unlikely
to explain the positive correlation between N~V/C~IV and Si~IV/C~IV observed
in AE~Aqr and possibly BY~Cam (see Figs.~7 and 8). The observed correlation
between these ratios lends support to the hypothesis that confusion between
the N~V 2s--2p doublet and the Mg~II 3s--4p doublet, together with a low value
of the ionization parameter and hence a large value of the Si~IV/C~IV ratio,
is responsible for the apparent anomalous line ratios of BY~Cam and AE~Aqr.
However, none of the models reproduce the nearly perfect linear proportionality
between N~V/C~IV and Si~IV/C~IV observed in AE~Aqr.

There are several possibilities why the models and the observed line ratios
are discrepant. First, it is possible that we have failed to consider ionizing
spectra of the right type. Since the 10 eV blackbody appears to come closest to
providing agreement with the He~II/C~IV and Si~IV/C~IV ratios simultaneously,
it is possible that there are confusing lines which we have not included in
our models, or which become important under other conditions, which affect the
results. Alternatively, the emission region may consist of multiple components
with differing physical conditions. If so, the various components must have line
ratios which bracket the observed values. This is ruled out by our models: no
single set of models brackets the observed ratio of He~II/C~IV, for example. The
observed values of this ratio are bracketed by the photoionization models on
the high side and the coronal models on the low side, so that a superposition
of these models might provide consistent line ratios. However, we consider this
possibility to be somewhat contrived, and a more detailed exploration is needed
to test whether it can account for all the ratios simultaneously. Another
possible explanation for the He~II/C~IV ratio is that our assumption of a
stationary steady state is invalid for the line-emitting region. If, for
example, the heating and ionization of the gas occurs as the result of many
impulsive events, then the time-average value of the ratios could differ
significantly from the steady-state model predictions owing to the differing
timescales for relaxation of the upper levels of the He~II line from that of
C~IV. Such a scenario has been suggested to account for the strength of the
He~II lines from the Sun (\cite{ray90}). The differential behavior of the line
ratios from a given object could be due at least in part to changes in the
ionizing spectrum or ionization mechanism, rather than simply due to changes
solely in ionization parameter. This could account for the departures from the
variability behavior predicted by the models. In spite of the difficulty in
reproducing the observed line ratios, steady-state photoionization models are
capable of fitting the absolute strengths of the observed lines (Ko et~al.\
1996). 

\section{Summary}

We have presented a statistical analysis of the Si~IV/C~IV, N~V/C~IV, and 
He~II/C~IV emission line ratios of 20 CVs based on $\approx 430$ UV spectra
extracted from the {\it IUE\/} ULDA. We find for most systems that these ratios
are clustered within a range of $\sim 1$ decade for log Si~IV/C~IV $\approx
-0.5$ and log He~II/C~IV $\approx -1.0$ and $\sim 1.5$ decades for log N~V/C~IV
$\approx -0.25$. The larger range of the N~V/C~IV ratio is due largely to the
large line ratios of the eclipsing dwarf novae Z~Cha and OY~Car; otherwise, the
range of log N~V/C~IV is $\sim 1$ decade centered on $\approx -0.5$. The
clearest difference between magnetic and non-magnetic CVs is the He~II/C~IV
ratio, which is $\sim 0.25$ decades larger in magnetic systems.

To place constraints on the excitation mechanism and the physical conditions
of the line-emitting gas of CVs, we have investigated the theoretical line
ratios of gas in either photoionization and collisional ionization equilibrium.
Given the uncertain and variable geometry, density, optical depth of the
line-emitting gas and the shape and luminosity of the ionizing spectrum, we
have restricted ourselves to consideration of simple slab models each with fixed
gas composition, density, and column density. The variables have been the shape
and ionization parameter of the ionizing spectrum and the density and column
density of the slab; for the collisional models, the temperature was varied.
Line emission is produced in these models by recombination or collisional
excitation by thermal electrons or by excitation by the ionizing continuum
(``scattering'').

Within the confines of these simple models, we find little agreement between the
observations and any of the models. Specifically, the observed Si~IV/C~IV line
ratios are reproduced by many of the models, but the predicted N~V/C~IV line
ratios are simultaneously too low by typically $\sim 0.5$ decades. Worse, for no
parameters are any of the models able to reproduce the observed He~II/C~IV line
ratios; this ratio is far too small in the collisional and scattering models and
too large by typically $\sim 0.5$ decades in the photoionization models. Among
the latter, the 10 eV blackbody models do the best job of reproducing the
three line ratios simultaneously, but the match with the N~V/C~IV line ratio is
accomplished only if the observed emission feature near 1240~\AA \ is due to the
Mg~II 3s--4p doublet at 1239.93, 1240.39~\AA \ instead of the N~V 2s--2p doublet
at 1238.82, 1242.80~\AA . 

Despite the above generally unfavorable comparisons between observations and
simple photoionization and collisional models, our investigation has proven
useful in revealing both the problems and promises of understanding the UV line
ratios of CVs. Future detailed work could be profitably performed on any and
all of the above CV subclasses with more detail in the  shape of the ionizing
spectrum and the geometrical distribution, density, and column density of the
emission region(s). Where the distance is well known, not only line ratios but
absolute line strengths can be fit. With larger effective area, weaker lines,
less subject to optical depth effects, can be included. With higher spectral
resolution, the UV lines be can uniquely identified, thus removing the annoying
ambiguity of some of the line identifications. Additional constraints on the
physical conditions and optical depth of the line-emitting gas is possible if
the UV doublets are resolved. At comparable or slightly higher spectral higher
resolution, the velocity field of the line-emitting gas can be constrained to
constrain the ionization parameter and hence the density. With realistic photon
transport in the models, the line shapes further constrain the models. By
extending the bandpass into the far-UV, lines from species with both lower
and higher ionization potentials (e.g., C~III, N~III, O~VI, P~V, S~IV, S~VI)
provide additional diagnostics. The UV data can and is being obtained with
{\it HST\/}, but to obtain the far-UV data, we require the likes of {\it HUT\/}
(e.g., \cite{lon96}), {\it ORFEUS\/} (e.g., \cite{ray95}), and {\it FUSE\/}.

\acknowledgments

We thank John Raymond for useful insights and suggestions and the referee for
helpful comments which significantly improved the original manuscript. Work at
Lawrence Livermore National Laboratory was performed under the auspices of the
U.S.\ Department of Energy under contract No.~W-7405-Eng-48. 

\clearpage    


\begin{center}
\begin{tabular}{cccc}
\multicolumn{4}{c}{\bf TABLE 1}\\
\multicolumn{4}{c}{PHOTOIONIZATION MODEL LINE RATIOS}\\
\tableline
\tableline
$\log\xi $& log Si~IV/C~IV& log N~V/C~IV& log He~II/C~IV\\
\tableline 
\multicolumn{4}{l}{Model A:
                   30 eV Blackbody Spectrum: 
                  \vrule height6pt width5pt depth-1pt}\\
\tableline 
   $-1.5$&     $?0.00$&     $-2.02$&     $?0.68$\\
   $-1.0$&     $-0.55$&     $-1.40$&     $-0.44$\\
   $-0.5$&     $-1.12$&     $-0.97$&     $-0.90$\\
   $?0.0$&     $-1.56$&     $-0.59$&     $-0.76$\\
   $?0.5$&     $-2.20$&     $-0.37$&     $-0.58$\\
   $?1.0$&     $-2.92$&     $-0.24$&     $-0.41$\\
\tableline
\multicolumn{4}{l}{Model B:
                   10 keV Bremsstrahlung Spectrum:
                   $\bullet $}\\
\tableline
   $-1.5$&     $?0.78$&     $?0.91$&     $?3.65$\\
   $-1.0$&     $?0.73$&     $-0.15$&     $?2.99$\\
   $-0.5$&     $?0.62$&     $-1.38$&     $?1.87$\\
   $?0.0$&     $?0.19$&     $-1.39$&     $?0.52$\\
   $?0.5$&     $-0.76$&     $-1.15$&     $-0.64$\\
   $?1.0$&     $-1.05$&     $-0.68$&     $-0.84$\\
\tableline
\multicolumn{4}{l}{Model C:
                   50 eV Blackbody Spectrum:
                   $\times $}\\
\tableline
   $-3.0$&     $?1.54$&     $?2.18$&     $?5.06$\\
   $-2.5$&     $?1.26$&     $?1.04$&     $?4.19$\\
   $-2.0$&     $?0.97$&     $-0.25$&     $?3.16$\\
   $-1.5$&     $?0.67$&     $-1.66$&     $?1.90$\\
   $-1.0$&     $?0.11$&     $-1.69$&     $?0.47$\\
   $-0.5$&     $-1.00$&     $-1.17$&     $-0.71$\\
   $?0.0$&     $-2.01$&     $-0.97$&     $-0.90$\\
   $?0.5$&     $-2.69$&     $-0.62$&     $-0.67$\\
   $?1.0$&     $-3.20$&     $-0.26$&     $-0.32$\\
   $?1.5$&     $-3.56$&     $?0.22$&     $?0.31$\\
\tableline
\end{tabular}
\end{center}

\clearpage    


\begin{center}
\begin{tabular}{cccc}
\multicolumn{4}{c}{\bf TABLE 1 --- continued}\\
\multicolumn{4}{c}{PHOTOIONIZATION MODEL LINE RATIOS}\\
\tableline
\tableline
$\log\xi $& log Si~IV/C~IV& log N~V/C~IV& log He~II/C~IV\\
\tableline 
\multicolumn{4}{l}{Model D:
                   10 eV Blackbody Spectrum:
                   +}\\
\tableline
   $-3.0$&     $-1.86$&     $?0.35$&     $?3.50$\\
   $-2.5$&     $-1.50$&     $-0.90$&     $?2.30$\\
   $-2.0$&     $-0.86$&     $-2.23$&     $?1.06$\\
   $-1.5$&     $-0.48$&     $-2.90$&     $?0.15$\\
   $-1.0$&     $-0.71$&     $-2.52$&     $-0.48$\\
   $-0.5$&     $-0.90$&     $-1.88$&     $-0.75$\\
   $?0.0$&     $-0.80$&     $-1.08$&     $-0.79$\\
   $?0.5$&     $-0.58$&     $-0.31$&     $-0.69$\\
   $?1.0$&     $-0.45$&     $?0.25$&     $-0.48$\\
   $?1.5$&     $-0.45$&     $?0.57$&     $-0.16$\\
\tableline 
\multicolumn{4}{l}{Model E:
                   30 eV Blackbody Spectrum, $\log N_{\rm H}=23$:
                   $\Box $}\\
\tableline
   $-1.5$&     $?0.15$&     $-1.83$&     $?1.46$\\
   $-1.0$&     $-0.31$&     $-1.47$&     $?0.18$\\
   $-0.5$&     $-0.74$&     $-1.13$&     $-0.49$\\
   $?0.0$&     $-0.90$&     $-0.96$&     $-0.59$\\
   $?0.5$&     $-0.93$&     $-0.92$&     $-0.58$\\
   $?1.0$&     $-0.93$&     $-0.91$&     $-0.58$\\
   $?1.5$&     $-0.93$&     $-0.89$&     $-0.58$\\
\tableline
\multicolumn{4}{l}{Model F:
                   10 keV Bremsstrahlung Spectrum, $\log N_{\rm H}=23$:
                   $\circ $}\\
\tableline
   $-1.5$&     $?0.86$&     $?1.33$&     $?4.20$\\
   $-1.0$&     $?0.77$&     $?0.44$&     $?3.49$\\
   $-0.5$&     $?0.67$&     $-0.81$&     $?2.42$\\
   $?0.0$&     $?0.30$&     $-1.36$&     $?1.05$\\
   $?0.5$&     $-0.37$&     $-1.15$&     $-0.07$\\
   $?1.0$&     $-0.69$&     $-0.91$&     $-0.44$\\
   $?1.5$&     $-0.73$&     $-0.79$&     $-0.48$\\
\tableline
\end{tabular}
\end{center}

\clearpage    


\begin{center}
\begin{tabular}{cccc}
\multicolumn{4}{c}{\bf TABLE 2}\\
\multicolumn{4}{c}{SCATTERING MODEL LINE RATIOS}\\
\tableline
\tableline
$\log\xi $& log Si~IV/C~IV& log N~V/C~IV& log He~II/C~IV\\
\tableline 
\multicolumn{4}{l}{Model A$^\prime $:
                   30 eV Blackbody Spectrum:
                  \vrule height6pt width5pt depth-1pt}\\
\tableline 
   $-1.5$&    $?0.36$&     $-1.44$&     $*-9.44$\\
   $-1.0$&    $-0.26$&     $-0.98$&     $*-8.84$\\
   $-0.5$&    $-1.06$&     $-0.65$&     $*-8.04$\\
   $?0.0$&    $-1.86$&     $-0.33$&     $*-6.40$\\
   $?0.5$&    $-3.09$&     $-0.19$&     $*-4.86$\\
   $?1.0$&    $-4.65$&     $-0.13$&     $*-3.35$\\
   $?1.5$&    $-6.36$&     $-0.10$&     $*-2.23$\\
\tableline
\multicolumn{4}{l}{Model B$^\prime $:
                   10 keV Bremsstrahlung Spectrum:
                   $\bullet $}\\
\tableline
   $-1.5$&    $?1.26$&     $?0.68$&     $-12.02$\\
   $-1.0$&    $?1.20$&     $?0.08$&     $-10.71$\\
   $-0.5$&    $?1.07$&     $-0.71$&     $*-9.49$\\
   $?0.0$&    $?0.68$&     $-0.86$&     $*-8.10$\\
   $?0.5$&    $-0.59$&     $-0.81$&     $*-7.14$\\
   $?1.0$&    $-2.25$&     $-0.47$&     $*-4.57$\\
   $?1.5$&    $-5.26$&     $-0.06$&     $*-0.79$\\
\tableline
\multicolumn{4}{l}{Model C$^\prime $:
                   50 eV Blackbody Spectrum:
                   $\times $}\\
\tableline
   $-3.0$&    $?2.01$&     $?1.86$&     $-11.36$\\
   $-2.5$&    $?1.70$&     $?1.02$&     $-10.81$\\
   $-2.0$&    $?1.41$&     $?0.07$&     $-10.32$\\
   $-1.5$&    $?1.10$&     $-0.97$&     $*-9.69$\\
   $-1.0$&    $?0.53$&     $-1.24$&     $*-8.85$\\
   $-0.5$&    $-0.75$&     $-0.79$&     $*-8.45$\\
   $?0.0$&    $-2.23$&     $-0.68$&     $*-7.13$\\
   $?0.5$&    $-4.35$&     $-0.42$&     $*-5.09$\\
   $?1.0$&    $-6.98$&     $-0.14$&     $*-3.19$\\
   $?1.5$&    $-\infty$&   $?0.26$&     $*-1.42$\\
\tableline
\end{tabular}
\end{center}

\clearpage    


\begin{center}
\begin{tabular}{cccc}
\multicolumn{4}{c}{\bf TABLE 2 --- continued}\\
\multicolumn{4}{c}{SCATTERING MODEL LINE RATIOS}\\
\tableline
\tableline
$\log\xi $& log Si~IV/C~IV& log N~V/C~IV& log He~II/C~IV\\
\tableline
\multicolumn{4}{l}{Model D$^\prime $:
                   10 eV Blackbody Spectrum:
                   +}\\
\tableline
   $-3.0$&    $-1.28$&     $?0.35$&     $-11.04$\\
   $-2.5$&    $-1.24$&     $-0.48$&     $-10.32$\\
   $-2.0$&    $-0.61$&     $-1.41$&     $*-9.50$\\
   $-1.5$&    $-0.13$&     $-2.22$&     $*-9.62$\\
   $-1.0$&    $-0.45$&     $-2.06$&     $*-9.77$\\
   $-0.5$&    $-0.85$&     $-1.48$&     $*-9.57$\\
   $?0.0$&    $-1.08$&     $-0.74$&     $*-8.62$\\
   $?0.5$&    $-1.16$&     $-0.02$&     $*-7.52$\\
   $?1.0$&    $-1.19$&     $?0.51$&     $*-6.85$\\
   $?1.5$&    $-1.20$&     $?0.81$&     $*-6.52$\\
\tableline
\multicolumn{4}{l}{Model E$^\prime $:
                   30 eV Blackbody Spectrum, $\log N_{\rm H}=23$:
                   $\Box $}\\
\tableline
   $-1.5$&     $?2.01$&     $?0.92$&     $*-6.19$\\
   $-1.0$&     $?1.85$&     $?0.27$&     $*-5.06$\\
   $-0.5$&     $?1.85$&     $-0.03$&     $*-3.90$\\
   $?0.0$&     $?1.90$&     $-0.01$&     $*-2.24$\\
   $?0.5$&     $?1.90$&     $-0.02$&     $*-0.77$\\
   $?1.0$&     $?1.92$&     $-0.18$&     $*?0.58$\\
   $?1.5$&     $?1.96$&     $?0.09$&     $*?1.91$\\
\tableline
\multicolumn{4}{l}{Model F$^\prime $:
                   10 keV Bremsstrahlung Spectrum, $\log N_{\rm H}=23$:
                   $\circ $}\\
\tableline
   $-1.5$&     $?1.23$&     $?1.21$&     $-11.89$\\
   $-1.0$&     $?1.19$&     $?0.77$&     $-10.29$\\
   $-0.5$&     $?1.19$&     $?0.28$&     $*-8.47$\\
   $?0.0$&     $?1.19$&     $-0.21$&     $*-6.36$\\
   $?0.5$&     $?1.14$&     $-0.62$&     $*-4.50$\\
   $?1.0$&     $?1.08$&     $-0.79$&     $*-1.93$\\
   $?1.5$&     $?1.05$&     $-0.81$&     $*?1.19$\\
\tableline
\end{tabular}
\end{center}

\clearpage    


\begin{center}
\begin{tabular}{cccc}
\multicolumn{4}{c}{\bf TABLE 3}\\
\multicolumn{4}{c}{COLLISIONAL IONIZATION MODEL LINE RATIOS}\\
\tableline
\tableline
$T$ (10{,}000 K)& log Si~IV/C~IV& log N~V/C~IV& log He~II/C~IV\\
\tableline 
\multicolumn{4}{l}{Model G:
                   Constant Temperature}\\
\tableline 
   *3&     $-0.12$&     $-2.63$&     $-1.67$\\
   *5&     $-0.46$&     $-3.00$&     $-2.61$\\
   *7&     $-1.83$&     $-2.91$&     $-3.54$\\
   *9&     $-2.65$&     $-2.79$&     $-4.04$\\
   11&     $-2.80$&     $-2.16$&     $-3.99$\\
   13&     $-2.65$&     $-1.29$&     $-3.58$\\
   15&     $-2.44$&     $-0.52$&     $-3.13$\\
   17&     $-2.29$&     $?0.04$&     $-2.74$\\
   19&     $-2.17$&     $?0.38$&     $-2.42$\\
   21&     $-2.09$&     $?0.52$&     $-2.16$\\
\tableline
\end{tabular}
\end{center}

\clearpage    


\clearpage    


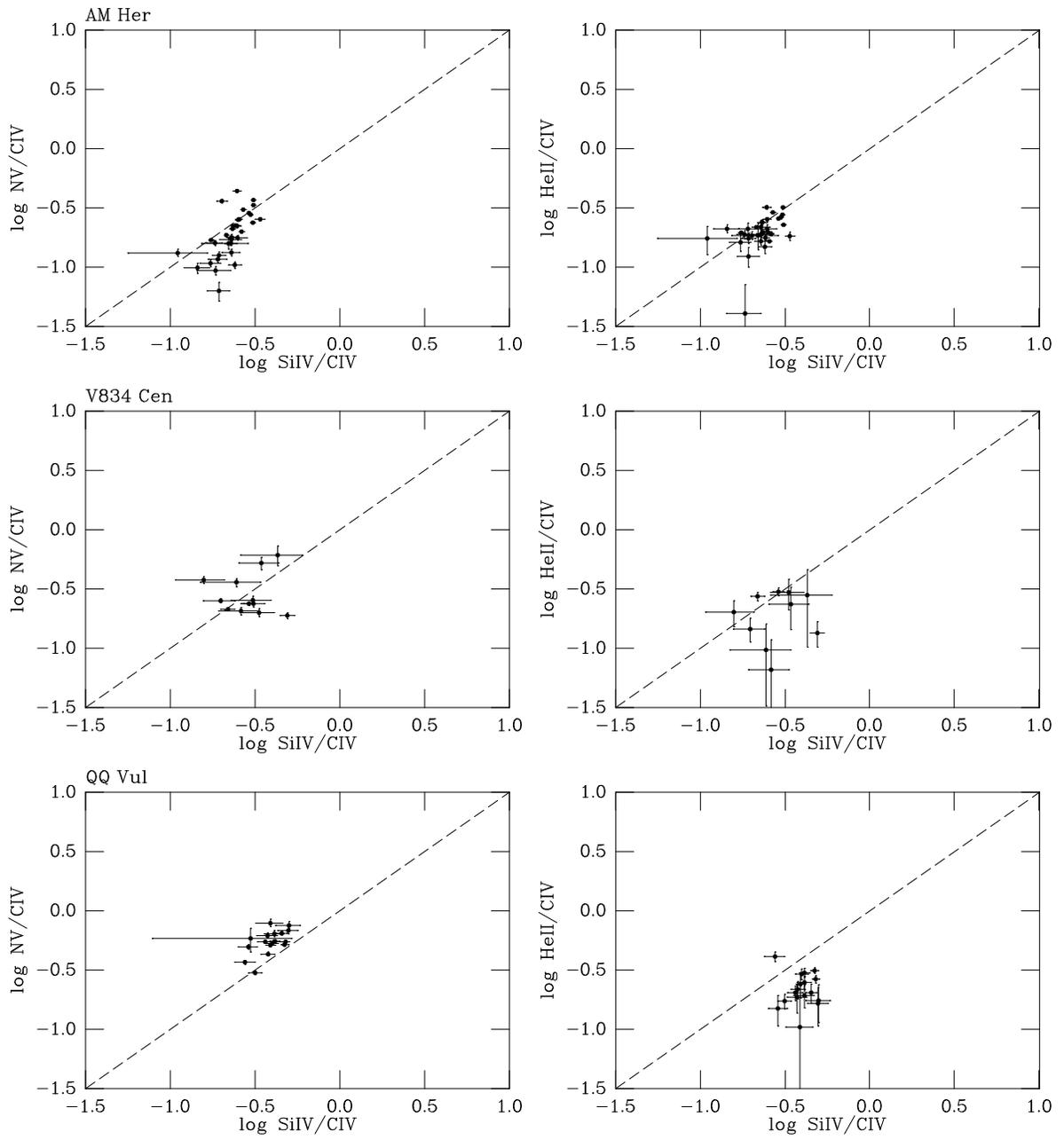
\begin{figure}
\caption{Line flux ratio diagrams for AM~Her stars.}
\end{figure}

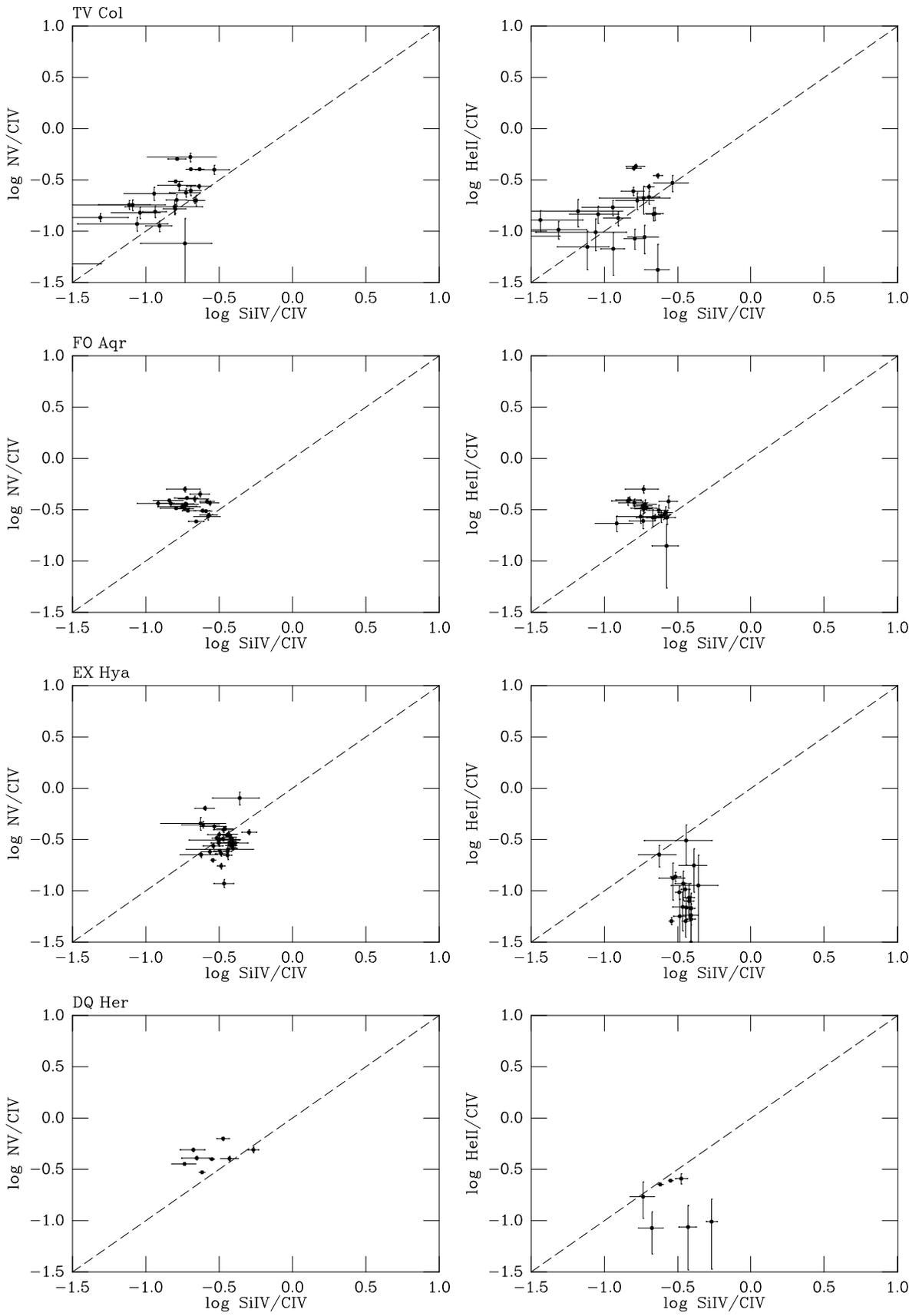
\begin{figure}
\caption{Line flux ratio diagrams for DQ~Her stars.}
\end{figure}

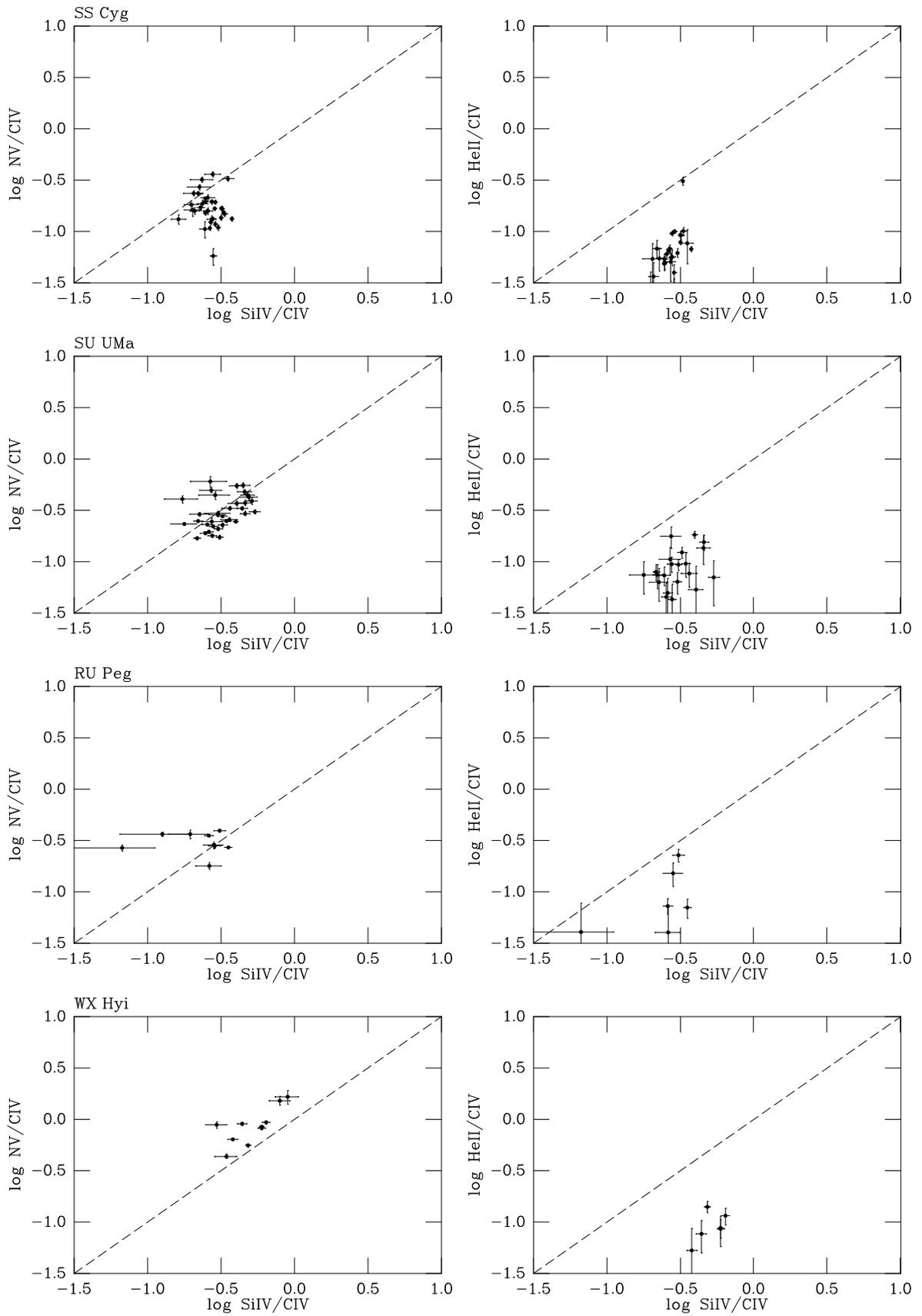
\begin{figure}
\caption{Line flux ratio diagrams for dwarf novae in quiescence.}
\end{figure}

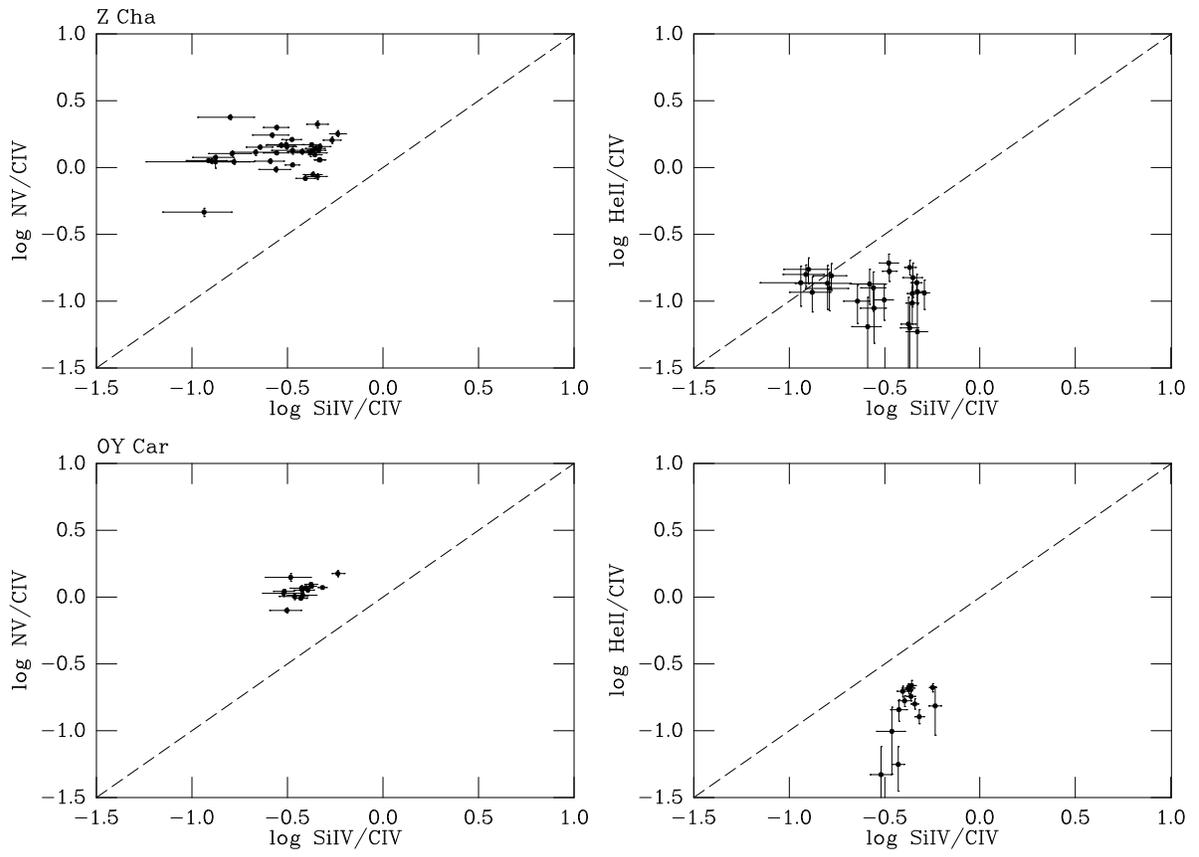
\begin{figure}
\caption{Line flux ratio diagrams for eclipsing dwarf novae.}
\end{figure}

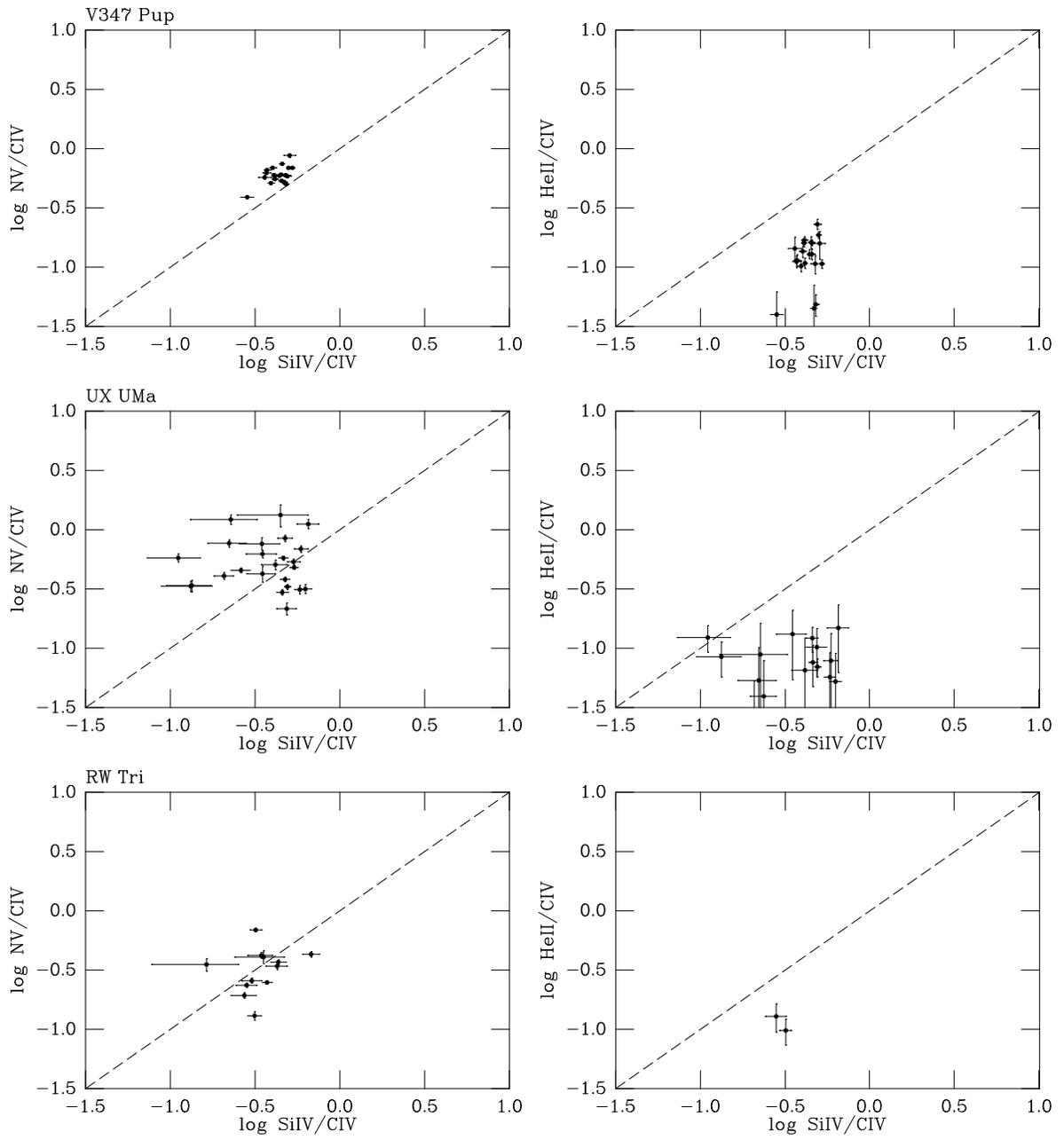
\begin{figure}
\caption{Line flux ratio diagrams for eclipsing nova-like variables.}
\end{figure}

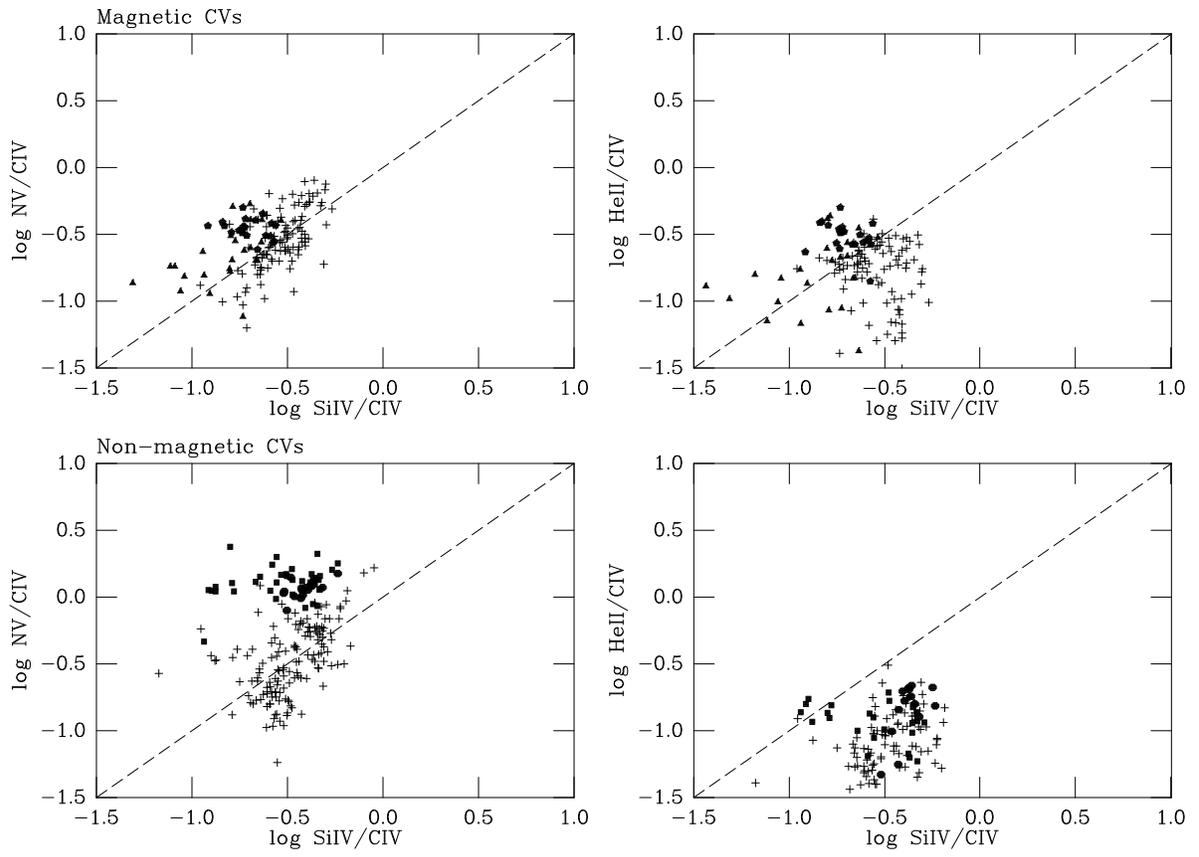
\begin{figure}
\caption{Line flux ratio diagrams for all ``normal'' magnetic and non-magnetic
         CVs. Error bars are suppressed to reduce clutter. Values for TV~Col 
         are shown as triangles, FO~Aqr as pentagons, Z~Cha as squares, OY~Car
	        as circles.}
\end{figure}

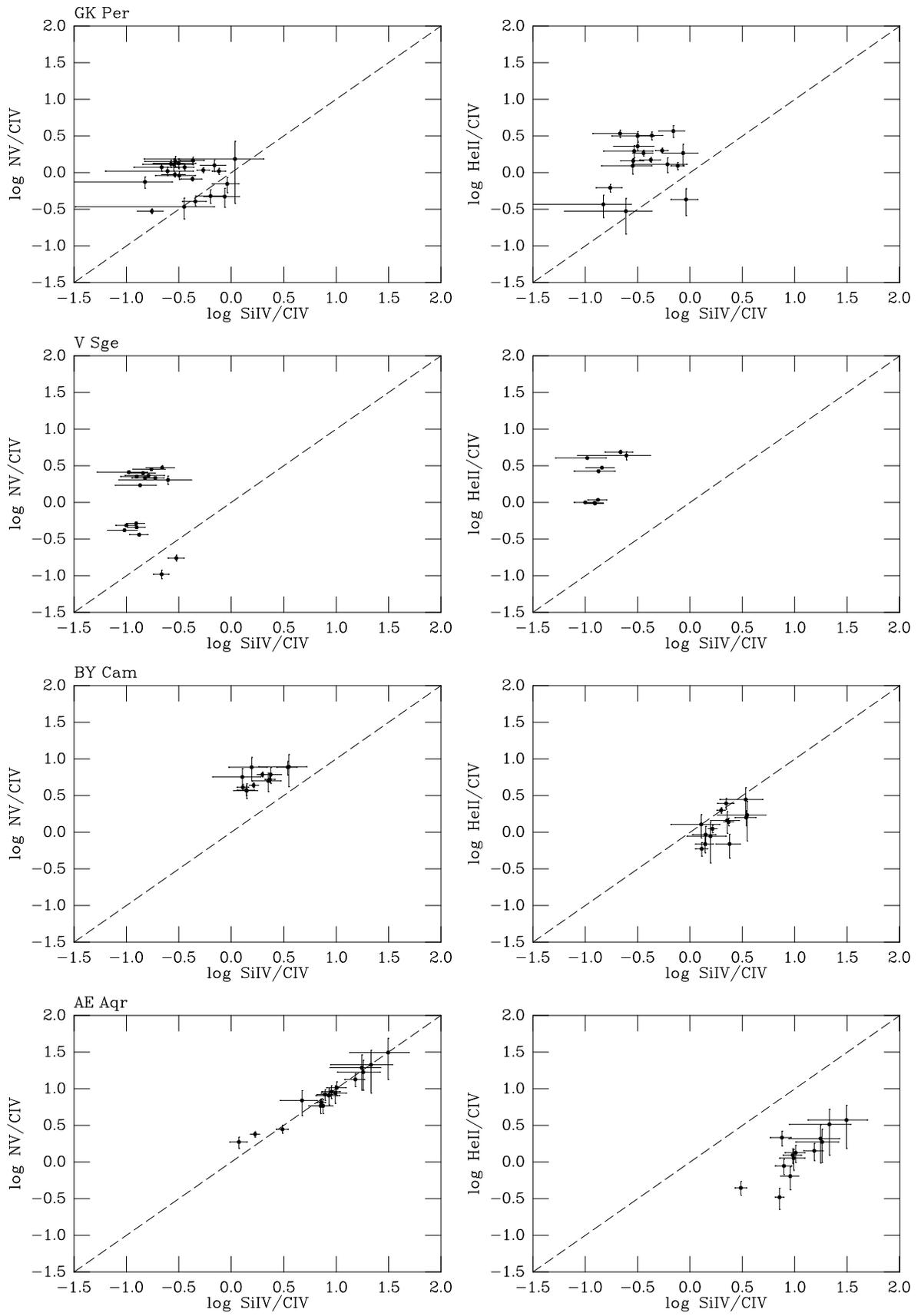
\begin{figure}
\caption{Line flux ratio diagrams for ``weird'' CVs. Note change of scale 
         relative to previous figures.}
\end{figure}

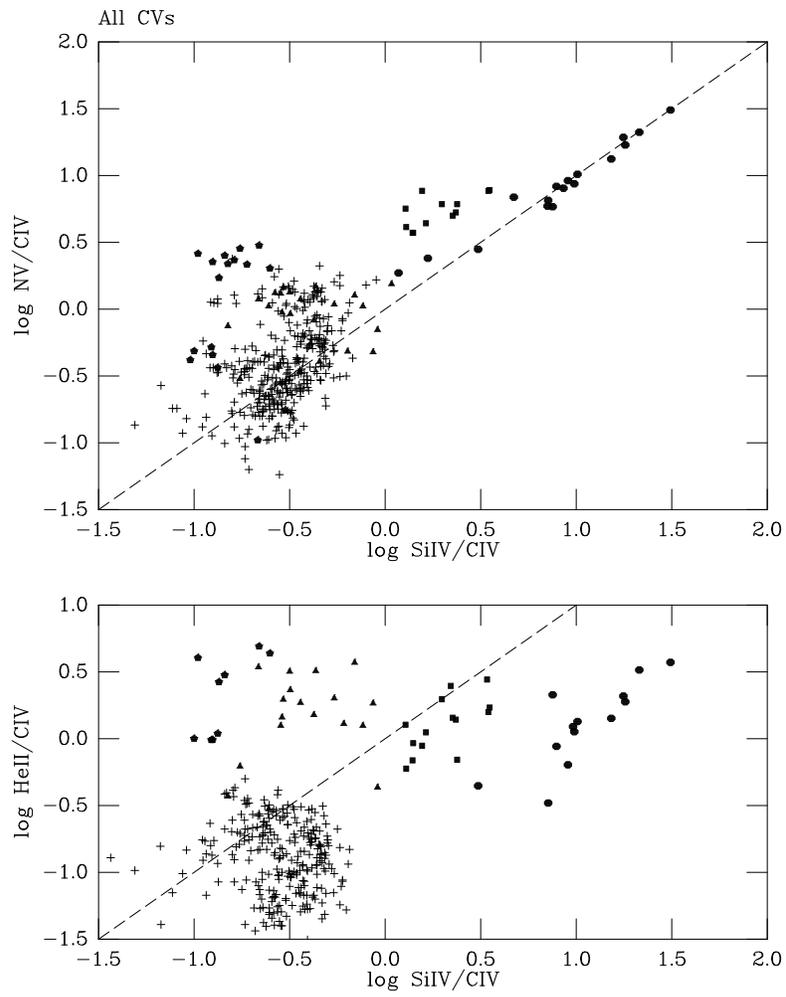
\begin{figure}
\caption{Line flux ratio diagram for all CVs. Error bars are suppressed to
         reduce clutter. Values for ``normal'' CVs are shown as crosses, 
         GK~Per as triangles, V~Sge as pentagons, BY~Cam as squares, AE~Aqr 
         as circles.}
\end{figure}

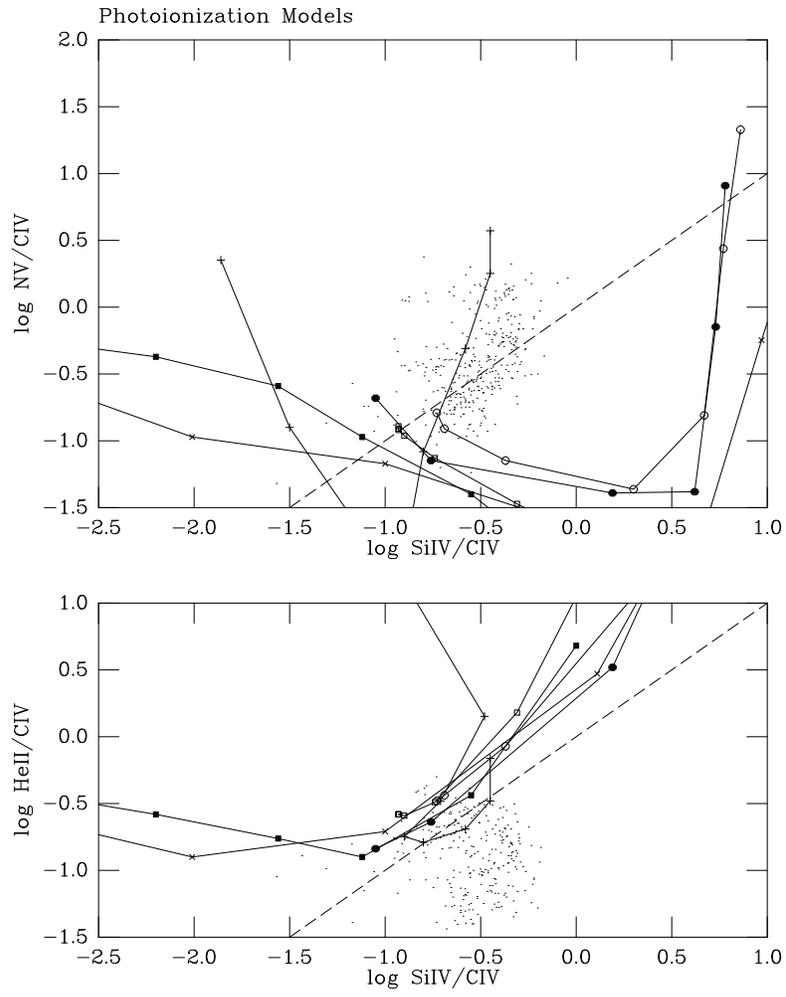
\begin{figure}
\caption{Model line flux ratio diagram for photoionized thermal emission
         models. Note that range of x axis differs from previous figure.
         Symbols are as follows:
         Model A = \vrule height6pt width5pt depth-1pt,
         Model B = $\bullet $,
         Model C = $\times $,
         Model D = +,
         Model E = $\Box $,
         Model F = $\circ $.
         Observed line flux ratios for all ``normal'' CVs are shown as points.}
\end{figure}

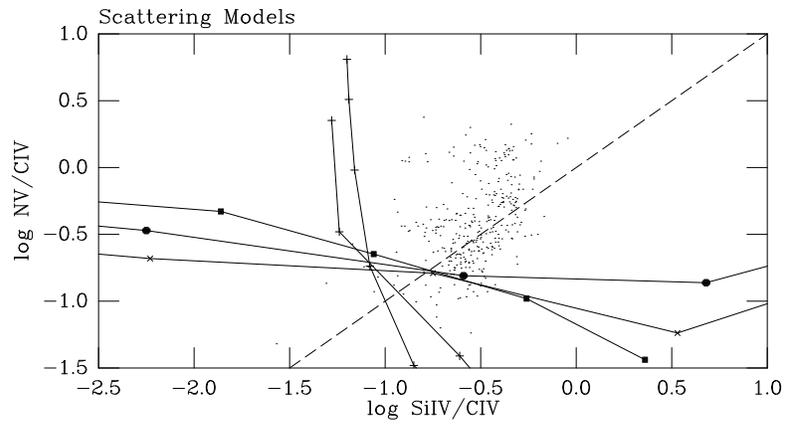
\begin{figure}
\caption{Model line flux ratio diagram for photoionized scattering line models. 
         Symbols are as follows:
         Model A$^\prime $ = \vrule height6pt width5pt depth-1pt,
         Model B$^\prime $ = $\bullet $,
         Model C$^\prime $ = $\times $,
         Model D$^\prime $ = +.
         Observed line flux ratios for all ``normal'' CVs are shown as points.}
\end{figure}

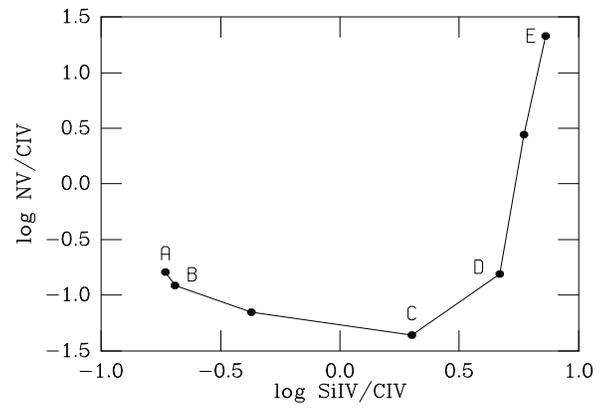
\begin{figure}
\caption{Schematic model line flux ratio diagram showing {\sf U}-shaped
         trajectory.}
\end{figure}

\end{document}